\documentclass[prd,aps,tightenlines,twocolumn,nofootinbib,preprintnumbers]{revtex4}
\usepackage{amsmath,amssymb}
\usepackage{slashed}

\newcommand{\be}{\begin{equation}}
\newcommand{\ee}{\end{equation}}
\newcommand{\bea}{\begin{eqnarray}}
\newcommand{\eea}{\end{eqnarray}}

\newcommand{\vo}{{\cal V}}

\begin{document}

\title{Inflation in string theory: a graceful exit to the real world}
\author{Michele Cicoli~$^1$, and Anupam Mazumdar~$^{2,~3}$}

\affiliation{$^1$~Deutsches Elektronen-Synchrotron DESY, Notkestrasse 85, 22603 Hamburg, Germany \\
$^2$~Physics Department, Lancaster University, Lancaster, LA1 4YB, United Kingdom\\
$^3$~Niels Bohr Institute, Copenhagen University, Blegdamsvej-17, DK-2100, Denmark}

%\date{\today}

\begin{abstract}
The most important criteria for a successful inflation are to explain the observed temperature anisotropy in the
cosmic microwave background radiation, and exiting inflation in a vacuum where it can excite the Standard Model
quarks and leptons required for the success of Big Bang Nucleosynthesis. In this paper we provide the first ever
closed string model of inflation where the inflaton couplings to hidden sector, moduli sector,
and visible sector fields can be computed, showing that inflation can lead to reheating the Standard Model
degrees of freedom before the electro-weak scale.
\end{abstract}

\preprint{DESY 10-166}

\maketitle

%%%%%%%%%%%%%%%%%%%%%%%%%%%%%%%%%%%%%%%%%%%%%%%%%%%%%%%%%%%%%%%%%%%%%%%%%%

\section{Introduction}

The inflationary paradigm has played a key r\^ole in explaining the large scale structure of the Universe
and the temperature anisotropy in the cosmic microwave background (CMB) radiation~\cite{WMAP7}.
It is believed that a scalar field, the inflaton, drives the inflationary dynamics.
During inflation the quantum fluctuations of the inflaton seed the initial perturbations
for the structure formation, and after inflation its coherent oscillations lead to reheat the Universe
with the observed light degrees of freedom of the Standard Model quarks and leptons
(for a review see~\cite{RM}).
These degrees of freedom need also to thermalise before Big Bang Nucleosynthesis (BBN)~\cite{BBN}.

In this regard visible sector models of inflation where inflation ends in the Standard Model {\it gauge invariant} vacuum,
are preferred with respect to the ones where the inflaton belongs to the hidden sector and couples to all particles in the model.
In the former case, there are very few models based on low scale supersymmetry (SUSY) as in the Minimal Supersymmetric Standard Model (MSSM),
where there exist {\it only} 2 inflaton candidates which carry Standard Model charges~\cite{AEGM},
and their decay populates the Universe with MSSM particles and dark matter~\cite{ADM}.
%However the lack of a supergravity embedding for these models, leads to assuming
%some features of the inflaton potential,
%like the inflaton couplings and the behaviour of higher order operators,
%which are crucial to check its flatness.

On the contrary there are many models of inflation where the inflaton is a Standard Model gauge singlet~\cite{RM}.
Promising examples are closed string inflationary models
in type IIB flux compactifications \cite{kahlerinfl, fiberinfl}.
The inflaton is a K\"{a}hler modulus parameterising the size of an internal cycle, and inflation
can be embedded within an UV complete theory where the $\eta$-problem can be solved
due to the no-scale structure of the potential and the fact that the volume mode of the internal 
manifold is kept stable during inflation.
Moreover the order of magnitude of all the inflaton couplings can be computed \cite{CM, astro}. 

Despite all these successes,
it is {\it still} hard to explain how the inflaton energy gets transferred primarily to visible degrees of freedom,
and not to hidden ones, since \textit{a priori a gauge singlet inflaton
has no preference to either the visible or the hidden sector}.

Thus the main challenge to build any hidden sector model of inflation is to reheat
the visible sector so that the thermal bath prior to BBN
is filled with the Standard Model hadrons. 

In order to fulfill this, the conditions are:
\begin{itemize}
\item The inflaton must decay primarily into the visible sector, and its coupling
to the hidden sector must be weak enough to prevent an overproduction of its degrees of freedom non-thermally
or non-perturbatively as in the case of preheating (for a review see~\cite{ABCM}).

\item The hidden sector must not contain very light species
since their presence at the time of BBN could modify the light-element abundance. The current LEP and BBN constraints
on extra light species is very tight, i.e. $\leq 4$~\cite{BBN}.

\item The visible sector must be sequestered from the hidden sector
so that the late decay of the hidden
degrees of freedom does not spoil the success of BBN or
overpopulate the dark matter abundance.
\end{itemize}
We stress that hidden sectors arise naturally in
string compactifications since they
come along with many internal cycles which have to be stabilised.
This is generically achieved wrapping
stacks of $Dp$-branes around $(p-3)$-cycles in order to generate perturbative and
non-perturbative effects that lead to moduli fixing. 
The presence of such $Dp$-branes is also in general needed to achieve tadpole cancellation 
which ensures the absence of anomalies in the 4D effective field theory.
Given that each stack of $D$-branes supports a
different field theory, the presence of hidden sectors turns out to be very generic in any such constructions.

In order to address all these issues related to the hidden sector, it is crucial to know the inflaton
couplings to all hidden and visible degrees of freedom. In this paper we shall present a type IIB closed string inflation model
where all the inflaton couplings can be derived within an UV complete theory \cite{CM}.
The knowledge of these couplings allows us to study the reheating of the visible degrees of freedom
which can be achieved with a temperature above the BBN temperature
(for other studies on (p)reheating in string cosmology see~\cite{Many-reheat,Preheating}).

The beauty of this {\it top-down} setup
is also that the flatness of the inflaton potential can be checked, and all the main phenomenological scales
can be generated with the following achievements:
\begin{itemize}
\item correct amount of CMB density perturbations;

\item right scale for grand unification theories (GUT);

\item TeV scale SUSY;

\item no cosmological moduli problem (CMP).
\end{itemize}

\section{Type IIB LARGE Volume Scenarios}
\label{2}

String compactifications typically give rise to a large number of moduli which can be ideal candidates to drive inflation.
It is crucial to lift these flat directions in order to determine the
features of the low energy effective field theory (like masses and coupling constants) and to avoid
the presence of unobserved long range fifth forces.

\subsection{Moduli stabilisation}

Moduli stabilisation is best understood in the context of type IIB where
background fluxes fix the dilaton and the
complex structure moduli at tree-level. On the other hand,
the stabilisation of the K\"{a}hler moduli requires to consider perturbative
and non-perturbative effects \cite{kklt, LVS, GenAnalofLVS}.
Expressing the K\"{a}hler moduli as $T_i=\tau_i+i b_i$, $i=1,...,h_{1,1}$, with $\tau_i$ the volume of an internal 4-cycle
$\Sigma_i$ and $b_i= \int_{\Sigma_i}C_4$, we shall focus on compactification manifolds whose volume reads
(with $\alpha>0,\gamma_i>0$ $\forall i$):
\be
\hat{\vo}=\alpha\left(\tau_1^{3/2}-\gamma_2\tau_2^{3/2}-\gamma_3\tau_3^{3/2}-\gamma_4\tau_4^{3/2}\right).
\ee
Assuming that the tadpole-cancelation condition can be satisfied by an appropriate choice
of background fluxes, we wrap a hidden sector $D7$-stack that undergoes gaugino-condensation
both around $\tau_2$ and $\tau_3$. This brane set-up induces a superpotential of the form
(the dilaton and the complex structure moduli are flux-stabilised at tree level and so they can be integrated out):
\be
W=W_0+A_2 e^{-a_2 T_2}+A_3 e^{-a_3 T_3}.
\ee
We focus on the case when the cycle $\tau_4$ supporting the visible sector shrinks down at the
singularity.
The visible sector is built via $D3$-branes at the quiver locus and the
gauge coupling is given by the dilaton $s$ (with $\langle s\rangle=g_s^{-1}$)
while $\tau_4$ enters as a flux-dependent correction: $4\pi g^{-2}=s+h(F)\tau_4$.
The K\"{a}hler potential with the leading order $\alpha'$ correction
can be expanded around $\tau_4=0$ \cite{quiver}:
\be
K=-2\ln\left(\vo+\frac{\xi s^{3/2}}{2}\right)+\lambda \frac{\tau_4^2}{\vo}-\ln\left(2 s\right),
\label{Kquivero}
\ee
with $\vo=\alpha\left(\tau_1^{3/2}-\gamma_2\tau_2^{3/2}-\gamma_3\tau_3^{3/2}\right)$.
In the absence of Standard Model singlets which can get a non-vanishing VEV,
an anomalous $U(1)$ on the visible sector cycle generates a D-term potential with a Fayet-Iliopoulos term:
\be
V_D=\left(\frac{2\pi}{s+h(F)\tau_4}\right)\xi_{FI}^2\text{ \ with \ }\xi_{FI}=\frac{Q_{\tau_4}\tau_4}{\vo},
\ee
while the leading order F-term potential reads (after minimising the axion directions):
\begin{eqnarray}
 V_F &=& \sum_{i=2}^3\frac{8 \, a_i^2 A_i^2}{3\alpha\gamma_i}
 \frac{\sqrt{\tau_i}e^{-2 a_i\tau_i}}{\vo}
 -4 W_0\sum_{i=2}^3\, a_i A_i \frac{\tau_i e^{-a_i\tau_i}}{\vo^2} \nonumber \\
 &&+\frac{3 \xi W_0^2}{4 g_s^{3/2}\vo^3}.
 \label{ygfdo}
\end{eqnarray}
The D-term potential scales as $V_D\sim \mathcal{O}(\vo^{-2})$,
while in the regime $\vo\sim e^{a_i\tau_i}$, $i=2,3$, $V_F\sim \mathcal{O}(\vo^{-3})$.
Hence at leading order $\xi_{FI}=0$ leads to $\tau_4\to 0$ fixing this cycle
at the quiver locus \cite{quiver}. On the other hand, $V_F$
completely stabilises $\tau_2$, $\tau_3$ and the
volume $\vo\simeq \alpha \tau_1^{3/2}$ at:
\be
a_i \langle \tau_i \rangle\,=\frac{1}{g_s}\left( \frac{\xi}{2
\alpha J} \right)^{2/3}\,\,\text{with}\,\,\,J=\sum_{i=2}^3\gamma_i/a_i^{3/2},
\label{minhea1}
\ee
and $\langle\vo\rangle = \left( \frac{ 3 \,\alpha \gamma_i }{4 a_i A_i}
 \right) W_0 \, \sqrt{\langle \tau_i\rangle }
 \; e^{a_i \langle \tau_i \rangle },\,\,\forall\, i=2,3$.

Moduli stabilisation is performed without fine tuning
the internal fluxes ($W_0\sim\mathcal{O}(1)$)
and the volume is fixed exponentially large in string units.
As a consequence, one has a very reliable effective field theory, as well as
a tool for the generation of phenomenologically desirable hierarchies.

\subsection{Particle physics phenomenology}

The particle phenomenology is governed by the background fluxes,
which break SUSY by the $F$-terms of the K\"{a}hler moduli and the dilaton
which then mediate this breaking to the visible sector.
However the $F$-term of $\tau_4$ vanishes since it is proportional
to $\xi_{FI}=0$: $F^4\sim e^{K/2} K^{4\bar{4}}W_0 \xi_{FI}= 0$.
Thus there is no local SUSY-breaking and the visible sector is sequestered,
implying that the soft terms can be suppressed with respect to $m_{3/2}$
by an inverse power of the volume. The main scales in the model are:
\begin{itemize}
\item GUT-scale: $M_{GUT}\sim {M_P}/{\vo^{1/3}}$,
\item String-scale: $M_s\sim {M_P}/{\vo^{1/2}}$, 
\item Kaluza-Klein scale: $M_{KK}\sim {M_P}/{\vo^{2/3}}$,
\item Gravitino mass: $m_{3/2}\sim {M_P}/{\vo}$,
\item Blow-up modes: $m_{\tau_i}\sim m_{3/2}$, $i=2,3$, 
\item Volume mode: $m_{\vo}\sim {M_P}/{\vo^{3/2}}$,
\item Soft-terms: $M_{soft}\sim {m_{3/2}^2}/{M_P}\sim {M_P}/{\vo^2}$.
\end{itemize}
Setting the volume $\vo \simeq 10^{6- 7}$ in string units, corresponding to
$M_s\simeq 10^{15}$ GeV, one can realise GUT theories, TeV scale SUSY and avoid any
CMP \cite{quiver}.

\subsection{Inflationary cosmology}

A very promising inflationary model can be embedded in this type IIB scenario
with the inflaton which is the size of the blow-up $\tau_2$ \cite{kahlerinfl}.
Displacing $\tau_2$ far from its minimum, due to the exponential suppression,
this field experiences a very flat direction which is suitable for inflation.
The other blow-up $\tau_3$, which sits
at its minimum while $\tau_2$ is slow rolling,
has been added to keep the volume stable during inflation.
In terms of the canonically normalised inflaton $\phi$, the potential looks like \cite{kahlerinfl}:
\be
V\simeq V_0 -\beta\left(\frac{\phi}{\vo}\right)^{4/3} e^{-a \vo^{2/3}\phi^{4/3}}.
\ee
This is a model of small field inflation, and so no detectable gravity waves are produced:
$r\equiv T/S \ll 1$. The spectral index is in good agreement with the observations:
$0.960<n_s<0.967$, and the requirement of generating enough density perturbations fixes $\vo \simeq 10^{6-7}$
which is the same value preferred by particle physics.

Potential problems come from $g_s$ corrections
to $K$ \cite{stringloops} which dominate the potential spoiling its flatness
once $\tau_2$ is displaced far from its minimum.
The only way-out is to fine-tune
these $g_s$ corrections small \cite{CM}.

In principle the non-perturbative potential for $\tau_2$ could also be generated by
a $D3$-brane instanton. In this way
hidden sectors and $g_s$ corrections would be absent.
However the requirement of having $\vo \simeq 10^{6-7}$ prevents this set-up
since both $\tau_2$ and $\tau_3$ would be fixed
smaller than the string scale where the effective field theory cannot be trusted anymore \cite{CM}.
Thus we realise that hidden sectors are always present in these models.

\subsection{Hidden sector configurations}

The hidden sectors on $\tau_i$, $i=2,3$, consist
in a supersymmetric field theory that undergoes gaugino condensation.
Broadly speaking we can entertain $3$ scenarios for the possible particle content and mass spectrum:
\begin{itemize}
\item The hidden sector is a pure $N=1$ supersymmetric Yang-Mills (SYM) theory
which due to strong dynamics confines in the IR at the scale $\Lambda$:
\be
\Lambda_i=M_s\, e^{- (4\pi g^{-2}) a_i/3},\,\,i=2,3.
\label{p1}
\ee
Given that $4\pi g^{-2}=\tau_i$,
and at the minimum $e^{-a_i\tau_i}\sim \vo^{-1}$, the order of magnitude of $\Lambda_i$ can be estimated as
$\Lambda_i\simeq M_P \vo^{-5/6}$.
The theory develops a mass gap and all particles acquire a mass of the order $\Lambda_i$
and are heavier than the inflaton after inflation since $m_{\tau_2}\simeq M_P/\vo<\Lambda_i$.
Thus the inflaton decay to hidden degrees of freedom is {\it kinematically forbidden}!

\item The hidden sector is a pure SYM theory plus a massless $U(1)$.
The mass-spectrum below $\Lambda$ consists of massless hidden photons and photini with
an $\mathcal{O}(M_{soft})$ mass due to SUSY-breaking effects.

\item The hidden sector is an $N=1$ $SU(N_c)$ theory with $N_f<(N_c-1)$ flavours.
The condensates of gauge bosons and gauginos get a mass of the order $\Lambda$
while all the matter condensates
get an $\mathcal{O}(M_{soft})$ mass due to SUSY-breaking effects except pion-like mesons
which remain massless in the presence of spontaneous chiral symmetry breaking.
If chiral symmetry is explicitly broken by a low energy Higgs-like mechanism,
all matter fields get a $\delta m \ll M_{soft}$ correction to their masses.
For an additional massless $U(1)$, there are also massless hidden photons and photini with
an $\mathcal{O}(M_{soft})$ mass.
\end{itemize}

\section{Reheating}

We shall now focus on the study of reheating of the MSSM degrees of freedom after the end of inflation.
In order to check what fraction of the inflaton energy is transferred
to hidden and visible degrees of freedom, we have to derive the moduli mass spectrum
and their couplings to all particles in the model.

\subsection{Canonical normalisation and mass spectrum}

The first step is to canonically normalise the moduli around the minimum
of their VEVs:
$\tau_i=\langle\tau_i\rangle+\delta \tau_i$, $\forall i$,
The fluctuations $\delta\tau_i$ are written in terms of the canonically normalised fields
$\delta\phi_i$ as $\delta\tau_i=\frac{1}{\sqrt{2}}C_{ij}\delta\phi_j$,
where $C_{ij}$ are the eigenvectors of the matrix
$\left(M^2\right)_{ij}\equiv \frac{1}{2}\left(K^{-1}\right)_{ik}V_{kj}$
whose eigenvalues $m_i^2$ are the moduli mass-squareds.

The form of $K$, eq. (\ref{Kquivero}), and $\langle\tau_4\rangle=0$
imply that at leading order $\tau_2$ does not mix with $\tau_4$.
However $\tau_2$ mixes with $s$
due to $\alpha'$ corrections to $K$ \cite{CM}:
\begin{eqnarray}
\delta \tau_1 &\sim&\mathcal{O}(\vo^{2/3})\delta \phi_1
+\sum_i\mathcal{O}(\vo^{1/6})\delta \phi_i,\,i=2,3,s, \notag \\
\delta \tau_i &\sim&\mathcal{O}(\vo^{1/2})\delta \phi_i+\mathcal{O}(1)\delta \phi_1
+\sum_j\mathcal{O}(\vo^{-1/2})\delta\phi_j,\,i=2,3, \notag \\
\delta \tau_4 &\sim&\mathcal{O}(\vo^{1/2})\delta \phi_4,
\label{cn2quiver} \\
\delta s &\sim&\mathcal{O}(1)\delta\phi_s+\mathcal{O}(\vo^{-1/2})\delta \phi_1+
\sum_i\mathcal{O}(\vo^{-1})\delta \phi_i,\,i=2,3, \notag
\end{eqnarray}
with $j=2,3,s$, $j\neq i$. The masses turn out to be:
\begin{equation} 
m^2_1\simeq \frac{M_P^2}{\vo^3}, \,\,\,m_i^2\simeq \frac{M_P^2}{\vo^2},
\forall i=2,3,s, \text{ \ and \ }m_4^2\simeq \frac{M_P^2}{\vo}.
\notag
\end{equation}

\subsection{Inflaton couplings}
\label{BI}

The inflaton coupling to visible and hidden degrees of freedom can be
derived from the moduli dependence of the kinetic and mass terms of open string modes.
The moduli are expanded around their VEVs and then expressed in terms of the
canonically normalised fields. This procedure led to the derivation of
the moduli couplings to all particles in the model \cite{CM, astro}, finding that
the strongest moduli decay rates are to hidden gauge bosons on $\tau_2$ and $\tau_3$,
and to visible gauge bosons at the $\tau_4$-singularity (see Tab. I).

\begin{table}[ht]
\begin{center}
\begin{tabular}{c||c|c|c|c|c}
  & $\delta\phi_1$ & $\delta\phi_2$
  & $\delta\phi_3$ & $\delta\phi_4$ & $\delta\phi_s$ \\
  \hline\hline
  \\ & & & & \vspace{-0.9cm}\\
  $(F_{\mu \nu}^{(2)} F^{\mu \nu}_{(2)})$
  & $\frac{1}{M_P}$
  & $\frac{\vo^{1/2}}{M_P}$
  & $\frac{1}{\vo^{1/2}M_P}$
  & -
  & $\frac{1}{\vo^{1/2}M_P}$ \\
  \hline
  \\ & & & & \vspace{-0.9cm}\\
  $(F_{\mu \nu}^{(3)} F^{\mu \nu}_{(3)})$
  & $\frac{1}{M_P}$
  & $\frac{1}{\vo^{1/2}M_P}$
  & $\frac{\vo^{1/2}}{M_P}$
  & -
  & $\frac{1}{\vo^{1/2}M_P}$ \\
  \hline
  \\ & & & & \vspace{-0.9cm}\\
  $(F_{\mu \nu}^{(4)} F^{\mu \nu}_{(4)})$
  & $\frac{1}{\vo^{1/2}M_P}$
  & $\frac{1}{\vo M_P}$
  & $\frac{1}{\vo M_P}$
  & $\frac{\vo^{1/2}}{M_P}$
  & $\frac{1}{M_P}$
\end{tabular}
\end{center}
\caption{Moduli couplings to all gauge bosons in the model.}
\end{table}

\subsection{Moduli dynamics after inflation and reheating}

At the end of inflation, due to the steepness of the potential,
the inflaton $\tau_2$, which acts like a homogeneous condensate,
stops oscillating coherently around its minimum just after 2-3 oscillations due to a very violent
non-perturbative production of $\delta\tau_2$ quanta \cite{Preheating}.
The production of other degrees of freedom at preheating is instead less efficient.

According to the second of equations (\ref{cn2quiver}), our Universe is mostly filled with $\delta\phi_2$
plus some $\delta\phi_1$ and fewer $\delta\phi_3$ and $\delta\phi_s$-particles.
Thus the energy density is dominated by $\delta\phi_2$
whose perturbative decay leads to reheating.
Denoting as $g$ the visible gauge bosons and as $X_2$ and $X_3$ the hidden ones,
the coupling of $\delta\phi_2$ to $X_2 X_2$ is stronger than the one
to $X_3 X_3$ which, in turn, is stronger than the one to $g g$.
This is due to the geometric separation between $\tau_2$ and $\tau_3$,
and the sequestering of the visible sector at the $\tau_4$-singularity.

Hence the first decays are $\delta\phi_i \to X_i X_i$, $i=2,3$ with decay
rate $\Gamma\sim M_P/\vo^2$.
Thus the inflaton dumps all its energy to hidden, instead of visible, degrees of freedom
without reheating the visible sector.
We stress that there is no direct coupling between hidden
and visible degrees of freedom since they correspond to
two open string sectors localised in different regions of the Calabi-Yau, and so the
reheating of the visible sector cannot occur via the decay of hidden
to visible degrees of freedom.
Hence the only way-out is to forbid the decay of $\delta\phi_2$ to any hidden particle.
This forces us to consider on both $\tau_2$ and $\tau_3$ a pure $N=1$ SYM theory that develops a mass gap, so that
the decay of $\delta\phi_2$ to $X_i X_i$, $i=2,3$ is kinematically forbidden.

Then the first decay is $\delta\phi_s\to gg$ with $\Gamma\sim M_P/\vo^3$ but without leading to reheating
since the energy density is dominated by
$\delta\phi_2$. Reheating occurs only later on when $\delta\phi_2$ decays to visible gauge bosons
with total decay rate $\Gamma_{\delta\phi_2\to gg}^{TOT}\simeq (\ln\vo)^3 M_P\vo^{-5}$ \cite{CM}.
At the same time $\delta\phi_3\to gg$ without giving rise to reheating since $\delta\phi_3$ is not dominating the energy density.

The maximal reheating temperature for the visible sector in the approximation of sudden thermalisation can be worked out
equating $\Gamma_{\delta\phi_2\to gg}^{TOT}$ to $H \simeq \left(T_{RH}^{max}\right)^2/M_P$ \cite{CM}:
\be
T_{RH}^{max}\simeq \sqrt{\Gamma_{\delta\phi_2\to gg}^{TOT}M_P}\simeq(\ln\vo)^{3/2}\frac{M_P}{\vo^{5/2}}.
\label{TRH3}
\ee
For $\vo\simeq 10^{6-7}$, we obtain $T_{RH}^{max}\simeq 10^{2-4}$ GeV which is
higher than $T_{BBN}\simeq 1$ MeV, and so
it does not create any problem if the matter-antimatter asymmetry
could be realised in a non-thermal/thermal way.
Later on $\delta\phi_1$ decays to visible degrees of freedom out of thermal equilibrium without
suffering from the CMP since its decays before BBN:
$T_{\delta\phi_1\to gg}\simeq M_P\vo^{-11/4}\sim 10^2$ GeV.

\section{Discussion: Interplay between global and local issues}

In this paper we did not address any of the issues related to the brane construction of the visible sector 
given that the actual details of the embedding of the Standard Model (or any generalisation thereof like 
the MSSM or GUT theories) into string theory 
are completely irrelevant for the study of reheating in the context of closed string inflationary models 
within the framework of type IIB compactifications.

In fact, type IIB Calabi-Yau flux compactifications are characterised by the fact that physics decouples 
into local (or brane) and global (or bulk) issues that can be consistently studied separately. 
Some examples of brane issues that depend only on the local brane construction 
are finding the right chiral spectrum, gauge group and Yukawa couplings, 
while issues like moduli stabilisation, supersymmetry breaking, 
inflation and reheating are purely global.

Regarding the study of the transfer of the inflaton energy density to the Standard Model degrees of freedom, 
we emphasize that the only thing which has to be considered is the inflaton dynamics after inflation 
with the non-perturbative (preheating) and perturbative (reheating) particle production. 
In particular, the study of reheating via the inflaton decay involves only the knowledge 
of the inflaton mass and couplings to all the degrees of freedom in the theory, 
and the computation of the overall scaling of these couplings does not depend at all 
on any local detail of the Standard Model brane construction.
 
In fact, it is the overall volume of the Calabi-Yau compactification 
which sets the order of magnitude of all these couplings 
whose computation requires only the knowledge of the moduli dependence of the gauge kinetic function, 
the K\"{a}hler potential and the superpotential, together with 
the form of the soft supersymmetry-breaking terms. All these are global or bulk issues.

\section{Conclusions}

In this paper we presented a model of closed-string slow-roll inflation embedded in 
type IIB string compactifications where we have full control over the inflaton dynamics after inflation
and the reheating of the visible sector degrees of freedom
regardless of the local Standard Model construction.

We tried to bring string inflation closer to `the real world'
describing how to excite the visible sector degrees of freedom in a Calabi-Yau compactification
with a robust moduli stabilisation mechanism that allows to check the solution of the $\eta$-problem
for inflation and compute the order of magnitude of the moduli mass spectrum and coupling to all particles of the theory.

This paper is mostly based on \cite{CM} but it singles out for the first time the best working model of closed string inflation
which shows many interesting phenomenological features like the generation of the correct amount of density perturbations,
a viable reheating of the visible sector degrees of freedom 
which can take place substantially before Big-Bang nucleosynthesis,
the absence of any cosmological moduli problem, the presence of TeV-scale supersymmetry and the right scale for grand unification theories.
We stress that all these achievements can be  reached without the need of any fine-tuning.

In this sense, this paper represents a step forward with respect to \cite{CM}
which was just a general analysis of reheating
with the discussion of several models,
and the pointing out of some problems that one has to solve to have a viable reheating model.

\subsection*{Acknowledgments}

The research of AM is partly supported by the
``UNIVERSENET'' (MRTN-CT-2006-035863).

%%%%%%%%%%%%%%%%%%%%%%%%%%%%%%%%%%%%%%%%%%%%%%

%%%%%%%%%%%%%%%%%%%%%%%%%%%%%%%%%%%%%%%%%%%%%%%%%%%%%%%%%%%%%%%%%%%%%%%%%

\end{document}